\documentclass{PoS}
\newcommand{\dkap}{\Delta\kappa^{\gamma}}

\newcommand{\lam}{\lambda^{\gamma}}

\title{Anomalous trilinear and quartic $WW\gamma$, $WW\gamma\gamma$, $ZZ\gamma$ 
and $ZZ\gamma\gamma$ couplings
in photon induced processes at the LHC}

\ShortTitle{Anomalous couplings at the LHC}

\author{\speaker{Christophe Royon}\\
        CEA/IRFU/Service de physique des particules, CEA/Saclay,  91191 Gif-sur-Yvette cedex, France\\
        E-mail: \email{royon@hep.saclay.cea.fr}}

\author{Emilien Chapon\\
        CEA/IRFU/Service de physique des particules, CEA/Saclay,  91191 Gif-sur-Yvette cedex, France\\
        E-mail: \email{emilien.chapon@cea.fr}}

\author{Oldrich Kepka\\
        CEA/IRFU/Service de physique des particules, CEA/Saclay,  91191 Gif-sur-Yvette cedex, France\\
        IPNP, Faculty of Mathematics and Physics,
        Charles University, Prague \\
	Center for Particle Physics, Institute of Physics, Academy of Science,
	Prague \\
        E-mail: \email{kepkao@fzu.cz}}

\abstract{We first report on possible measurements at the LHC using the first data and a
luminosity of 10 pb$^{-1}$ of $W$ and $Z$ pair production via two-photon
exchange. This measurement
allows in particular to increase the present sensitivity on $WW\gamma \gamma$ and
$ZZ\gamma \gamma$ quartic anomalous couplings 
from the LEP experiments by
almost three orders of magnitude. We also discuss the possible improvements on
quartic and trilinear anomalous couplings at high luminosity at the LHC using
new forward proton taggers to be installed at 220 and 420 m from the CMS or ATLAS
detectors.
}

\FullConference{The 2009 Europhysics Conference on High Energy Physics,\\
		 July 16 - 22 2009\\
		 Krakow, Poland}

\begin{document}

\section{Anomalous quartic $WW\gamma\gamma$ and $ZZ\gamma\gamma$ couplings at
LHC}

The non-abelian gauge nature of the Standard Model (SM)
predicts the existence of quartic couplings
$WW\gamma \gamma$
between the $W$ bosons and the photons which can be probed directly at the 
Large Hadron Collider (LHC) at CERN.
The quartic coupling to the $Z$ boson $ZZ\gamma \gamma$ is not present in the
SM. High energy colliders such as the incoming LHC are the natural place to look
for anomalous quartic couplings between the photon and the $W$ or $Z$ bosons.
The process we want to study at the LHC 
corresponds to $pp \rightarrow pWWp$. In this photon-induced process, the two
quasi-real 
photons interact through the exchange of a virtual $W$ or $Z$, leading to a 
pair of $W$s or $Z$s
in the final state. The advantage of this kind of events is that they are
extremly clean, there are two $W$s (or $Z$s) which can be detected in the ATLAS or CMS
central detectors while the intact proton leave undetected in the beam
pipe. 

To study the sensitivity on quartic anomalous couplings at the LHC,
we restrict ourselves to the implementation of the genuine
quartic anomalous 
$\gamma \gamma WW$ and $\gamma \gamma ZZ$ using the lowest dimension
operators possible in the Lagrangian ~\cite{us} performed in the FPMC
generator~\cite{fpmc}. The implemented survival probability is 0.9 for QED
events~\cite{survival}.

The current best 95\% confidence level (C.L.) limits on the parameters of quartic anomalous couplings were
determined by the OPAL Collaboration~\cite{opal} at LEP 
and are given in Table~\ref{table2}.

We study the sensitivity on quartic anomalous 
$a_0^W$ and $a_C^W$ couplings
using the first data to be taken at the LHC, and namely a luminosity of 10 (or
100) pb$^{-1}$ which can be accumulated in a couple of days or weeks at a
center-of-mass energy of 10 TeV. To simplify the study, we limit
ourselves to the leptonic decays of the $W$ pair to electrons or muons.
The signal is characterised by the presence of two high transverse
momentum ($p_T$) leptons
(electrons or muons) reconstructed in the ATLAS central detector, and the 
absence of any other reconstructed object or energy flow since the scattered
protons leave undetected in the beam pipe. 

Let us now discuss each background in turn implemented in FPMC (the survival
probability for double pomeron exchange processes is assumed to be 0.03 at LHC
energies~\cite{survival}). The non-diffractive
$W$ pair production is suppressed by requesting the presence of two leptons and
nothing else in the ATLAS detector since the inclusive background always shows
some hadronic activity in the calorimeter or in the forward part of
the tracking detector.  The dilepton production through photon 
exchange (QED process) is
suppressed by requesting the presence of a leading lepton with $p_T>160$ GeV,
and of missing energy greater than 20 GeV, which is natural when one requests the
presence of two $W$s. The pure SM $W$ pair background (without any anomalous
couplings) via photon exchange is small
since the value of the cross section
is low (62 fb) and it is further suppressed by requiring a reconstructed lepton
with $p_T>160$ GeV.
The diffractive production of dileptons or $W$ pairs via double pomeron
exchange (DPE) leads to a negligible background after the exclusivity
cut. After all cuts, the background is found to be negligible. 
The reach on the different anomalous
couplings is given in Table~\ref{table2}.

At higher luminosity, the forward proton detectors~\cite{det} located at 220 and 420 m
along the beam direction far away from the ATLAS or CMS interaction points are
nedeed to remove the pile up background using the timing detectors 
(the scattered protons can be originated from additional soft interactions with
respect to the hard interaction producing the $W$ or $Z$ pairs). With a
luminosity of about 100 fb$^{-1}$, the sensitivity on anomalous quartic
couplings can be improved further by two orders of magnitude. This is specially
interesting to test higgless models which predict the existence of quartic
anomalous couplings at one loop of the order of a few 10$^{-6}$ GeV$^{-2}$.

\begin{table}[htb]
\centerline{
   \begin{tabular}{|c||c|c|c|}
    \hline
    \raisebox{-1.5ex}[0pt][0pt]{Couplings} & 
    OPAL limits & 
    \multicolumn{2}{c|}{Sensitivity @ $\mathcal{L} = 10$ (100) pb$^{-1}$} \\
    &  \small[GeV$^{-2}$] & 5$\sigma$ & 95\% CL \\ 
    \hline
    $a_0^W/\Lambda^2$ & [-0.020, 0.020] & 2.2 10$^{-4}$ & 1.0 10$^{-4}$\\
                      &                 & (7.3 10$^{-5}$) & (3.3 10$^{-5}$)\\ \hline               
    $a_C^W/\Lambda^2$ & [-0.052, 0.037] & 5.9 10$^{-4}$ & 3.5 10$^{-4}$\\
                      &                 & (2.4 10$^{-4}$) & (1.1 10$^{-4}$)\\ \hline               
    $a_0^Z/\Lambda^2$ & [-0.007, 0.023] & 1.0 10$^{-3}$ & 5.2 10$^{-4}$\\
                      &                 & (3.7 10$^{-4}$) & (1.7 10$^{-4}$)\\ \hline               
    $a_C^Z/\Lambda^2$ & [-0.029, 0.029] & 3.0 10$^{-3}$ & 1.8 10$^{-3}$\\
                      &                 & (1.3 10$^{-3}$) & (5.9 10$^{-4}$)\\ \hline               
    \hline
   \end{tabular}
   }
\caption{Limits on anomalous coupling coming
from the LEP OPAL experiment. 
Sensitivity at low luminosity. The $5\sigma$ discovery potentials as well
as the 95\% C.L. limits are given for 10 pb$^{-1}$ and 100 pb$^{-1}$ in
parenthesis.}
\label{table2}
\end{table}

\begin{figure*}
\includegraphics[scale=0.37]{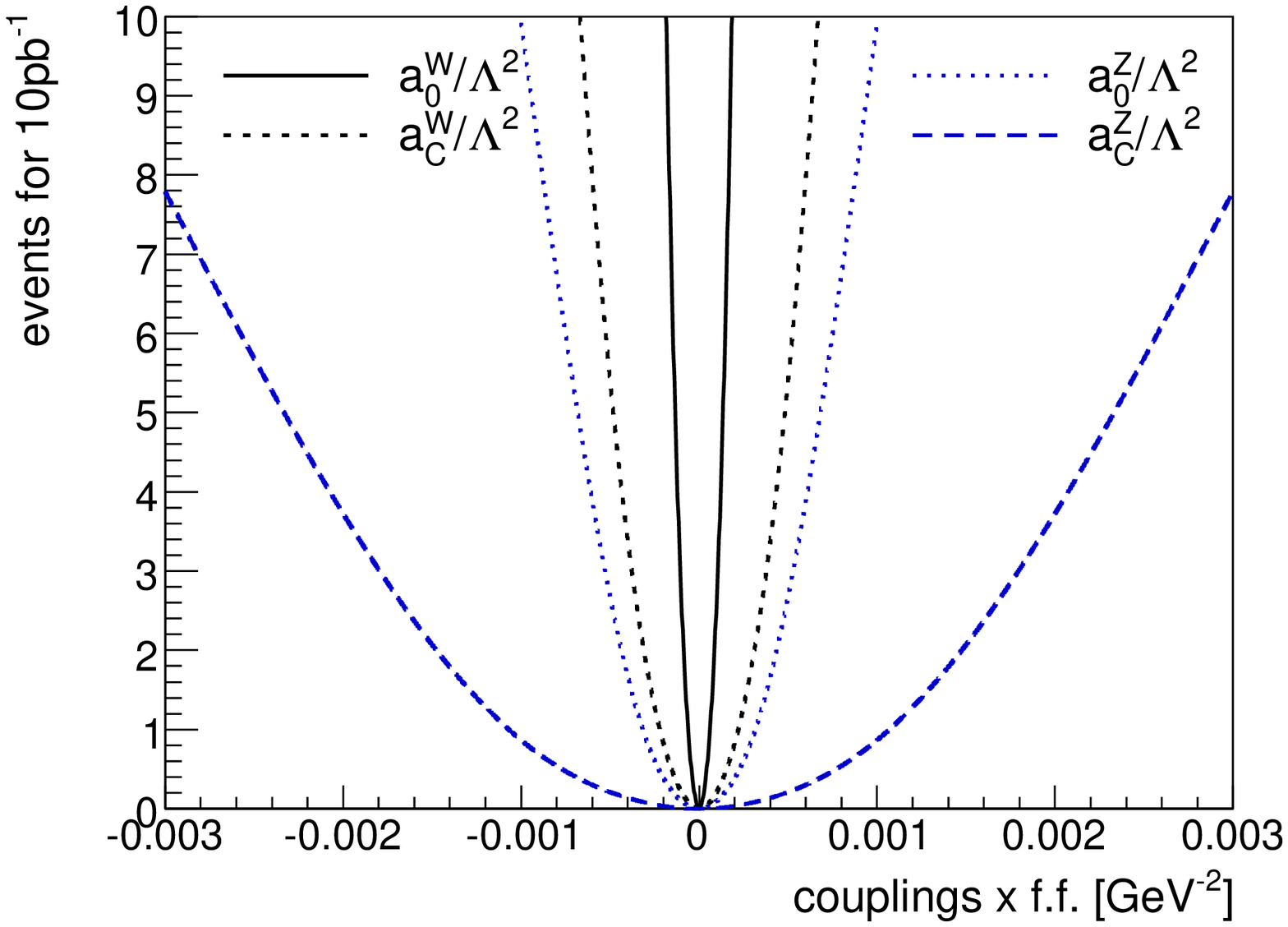}
\includegraphics[scale=0.37]{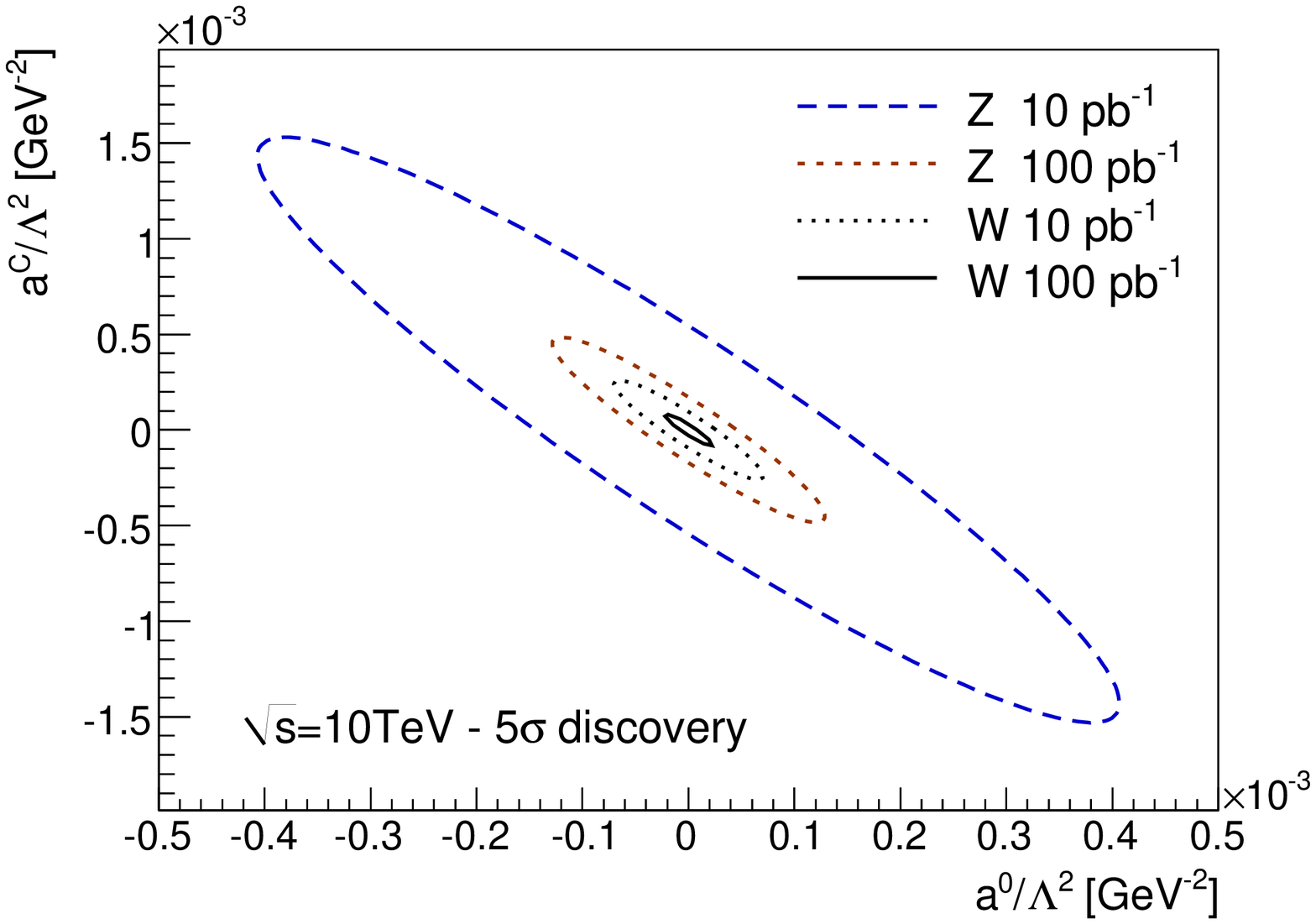}
\caption{\label{fig3} Left: Number of signal events after cuts as a function of the 
value of the quartic anomalous coupling for 10 pb$^{-1}$ at the LHC.
Right: $5\sigma$ discovery contours for the $WW$ and $ZZ$ quartic 
anomalous couplings at $\sqrt{s}=10$ TeV for luminosities of 10 and 
100 pb$^{-1}$.}
\end{figure*}

\section{Anomalous trilinear couplings at LHC}
At the LHC, it is also possible to study the anomalous trilinear couplings
$WW\gamma$ and $ZZ\gamma$. We consider the modification of the $WW\gamma$ 
triple gauge boson vertex with additional terms conserving $C-$ and $P-$
parity separately, that are parametrized with two anomalous parameters 
$\dkap$, $\lam$~\cite{olda}.
In order to obtain the best $S/\sqrt{B}$ ratio, the
$\xi$ acceptance was further optimized for the $\lam$ parameter.
The event is accepted if $\xi>0.05$.  
In case of $\dkap$, the full acceptance of the forward detectors is used since 
the difference between the enhanced and SM cross section
is almost flat around relevant values of the coupling $|\dkap|$. For 30 fb$^{-1}$,
the reach on  $\dkap$ and $\lam$ is respectively 0.043 and 0.034, improving the direct limits
from hadronic colliders by factors of 12 and 4 respectively (with respect to the LEP indirect 
limits, the improvement is only about a factor 2).
Using a luminosity of 200 fb$^{-1}$, present sensitivities
coming from the hadronic colliders
can be improved by about a factor 30, while the LEP sensitivity can be improved
by a factor 5. 

\begin{figure}
\begin{center}   
\includegraphics[scale=0.4]{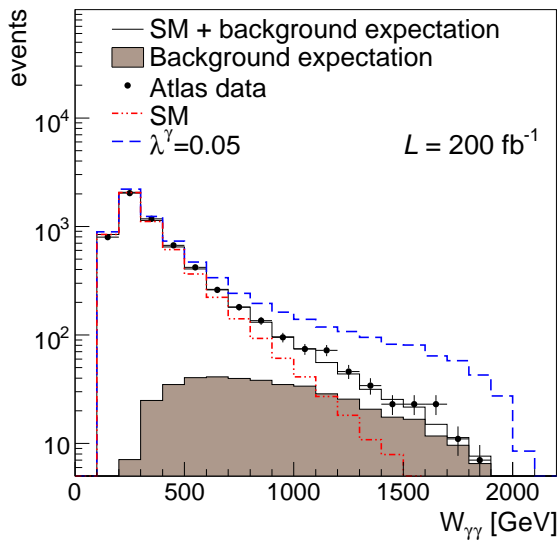}
\caption{Distributions of the $\gamma\gamma$ photon invariant mass $W_{\gamma\gamma}$ 
measured with the forward detectors using $W_{\gamma\gamma}=\sqrt{s\xi_1\xi_2}$. The 
effect of the $\lam$ anomalous parameter appears at high $\gamma\gamma$ 
invariant mass (dashed line).  The SM background is indicated in dot-dashed line, 
the DPE background as a shaded area and their combination
in full line. The black points show the ATLAS data smeared according to a Poisson distribution. }
\label{fig:Wspectrum200}
\end{center}
\end{figure}

It is worth noticing that many observed events are expected in the region $W_{\gamma\gamma}>1$ TeV where 
beyond standard model effects, such as SUSY, new strong dynamics at the TeV
scale, anomalous coupling, etc., are expected (see Fig.~\ref{fig:Wspectrum200}). 
It is expected that the LHC
experiments will collect 400 such events predicted by QED with $W>$1 TeV for a luminosity of 200
fb$^{-1}$ which will allow to probe further the SM expectations. 

\end{document}